\documentclass[journal,onecolumn]{IEEEtran}

\usepackage{amsmath,amssymb,amsfonts}
\usepackage[T1]{fontenc}
\usepackage[utf8]{inputenc}
\usepackage{graphicx}
\usepackage[export]{adjustbox}
\usepackage{booktabs}
\usepackage{multirow}
\usepackage{array}
\usepackage{float}
\usepackage{cite}
\usepackage{url}
\usepackage[unicode]{hyperref}
\graphicspath{{mathpix_assets/}}
\DeclareUnicodeCharacter{2713}{\checkmark}
\DeclareUnicodeCharacter{00D7}{\ensuremath{\times}}

\title{Neural Network-Based Impedance Identification and Stability Analysis for Double-Sided Feeding Railway Systems}
\author{Xiangyu Meng, Guiyang Hu, Zhigang Liu, Hui Wang, Guinan Zhang, Hongjian Lin, Mahdieh S. Sadabadi}

\begin{document}
\maketitle

\begin{abstract}
The double-sided power supply railway system increases the simultaneous operation of vehicles on the grid, potentially causing system instability and oscillation overvoltage issues. As vehicles frequently switch operating points during operation, it is essential to analyze system stability across a wide range of conditions. Therefore, accurately identifying the black-box impedance of vehicle converters at multiple operating points is crucial for studying railway vehicle-grid system stability. However, traditional impedance identification methods require extensive data and lack interpretability, leading to significant computational and data burdens. This study introduces an interpretable residual feedforward neural network (ResFNN) combined with SHapley Additive exPlanations for training vehicle impedance models, reducing data requirements while maintaining accuracy. Additionally, a component connection method is proposed for deriving the impedance matrix of a multivehicle railway system under the double-sided feeding mode. This method incorporates the dynamic mobility of vehicles and their positional distribution, and it utilizes the ResFNN to identify impedance for stability analysis. Real operational data from actual railway lines is used as case study to analyze the stability of the double-sided power supply railway system. The results demonstrate that this approach accurately assesses both lowfrequency and high-frequency instability issues.
\end{abstract}

Index Terms-Railway, vehicle, oscillation, stability, doublesided feeding, impedance identification.

\section*{Nomenclature}
\section*{ANN Artificial Neural Networks}
CCM Component Connection Method\\
GNSC Generalized Nyquist Stability Criterion\\
MSE Mean Squared Error\\
NARX Nonlinear Auto-regressive with eXogenous inputs\\
PP Parallel Point\\
ResFNN Residual Feedforward Neural Network\\
RNN Recurrent Neural Network\\
SHAP SHapley Additive exPlanations

\section*{I. Introduction}

THE demand for reliable, efficient, and high-capacity railway transportation is rapidly increasing. Traditional single-way power feeding modes in railway traction power systems suffer from neutral zones that can cause power interruptions and create barriers to power exchange [1]. However, the double-sided power feeding mode offers a more promising solution by eliminating neutral zones, thereby increasing power capacity and extending supply distance [2]. Despite the benefits of double-sided power feeding, the absence of neutral zones leads to electrical interconnection between multiple power supply sections. This interconnected environment results in a significant increase in the number of vehicles operating within the same power supply network. Current incident reports indicate that simultaneous operation of multiple vehicles in the grid can cause low-frequency oscillation (LFO) [3], harmonic instability (HIS) [4], and other instability issues [5], shown in Fig. 1. These factors can compromise system safety and stability. Given the heightened risk of instability associated with double-sided power feeding, it is critical to investigate the stability of the railway vehicle-grid system under this configuration.

The impedance-based analysis method is extensively employed to assess the stability of railway vehicle-grid system. This approach aligns with the EN 50388-2022 standard [5], which mandates that the real part of the vehicle input admittance remains positive within a specific frequency range to prevent harmonic instability. However, practical challenges hinder the creation of accurate vehicle impedance models. On the one hand, vehicle manufacturers are typically reluctant to disclose the internal control structures and parameters of their vehicles due to intellectual property concerns. This lack of transparency makes acquiring reliable vehicle impedance models virtually impossible. On the other hand, the intricate and nonlinear nature of modern railway vehicles poses a significant challenge to accurate modeling [6], [7]. Vehicles consist of numerous components, often with nonlinear interactions, further complicating the development of a comprehensive impedance model. To address the difficulties in constructing vehicle impedance models, researchers have focused on impedance identification methods. These methods are designed to estimate the input admittance or impedance characteristics of railway vehicles without needing direct access to proprietary internal details. By accurately identifying the impedance characteristics, it becomes feasible to analyze and enhance the stability of the railway vehicle-grid system under various power supply configurations.

Converter impedance identification methods can be broadly

\begin{figure}[H]
\begin{center}
  \includegraphics[alt={},max width=\columnwidth]{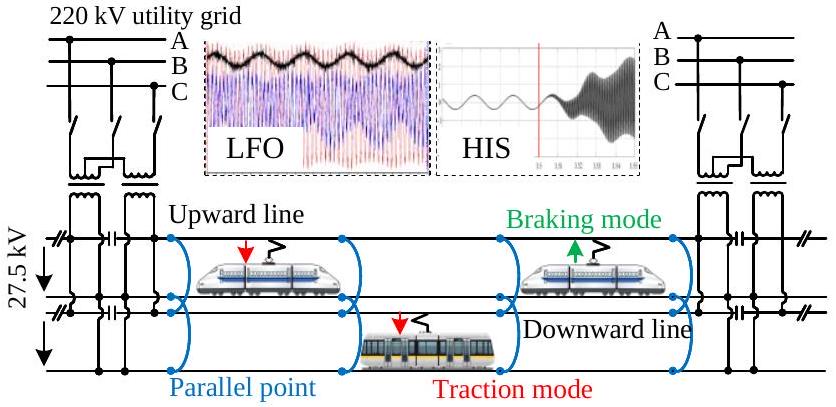}
\caption{Railway vehicle-grid system under all parallel double-sided direct power supply.}
\end{center}
\end{figure}

classified into two types: gray-box and black-box impedance identification [8]. The gray-box modeling method incorporates the prior knowledge of the system structure for further refinement. Notable gray-box models include the iterative least-squares technique [9] and the NARX model [10]. These models leverage the static behavior characteristics of converters, using known information to filter model structures and refine them with empirical data to accurately predict the dynamic behavior of power converters. However, graybox impedance identification requires prior knowledge of the converter's control or circuit structure, which is often unavailable due to intellectual property restrictions. In comparison, black-box modeling relies solely on input-output data to build models, without requiring any prior knowledge of the system's internal structure or physical characteristics [11]. Black-box methods can be further divided into linear and nonlinear categories. Linear models, such as polytopic model [12], [13], typically utilize transfer function models at a fixed operating point but are constrained by their limited range, as they only capture small signal behaviors and cannot encompass broader dynamic phenomena. Conversely, nonlinear black-box methods are more flexible, accommodating complex, larger signal variations.

Nonlinear black-box impedance identification methods aim to capture the nonlinear dynamic behavior of the system by leveraging large datasets to achieve highly accurate models. This has garnered significant attention in recent years. RNNs are used for time-domain modeling, and their results are later transformed for frequency-domain analysis [14]. However, these methods necessitate meticulous time-domain data design to effectively capture broad frequency-domain dynamics. As a result, model training and parameter configuration pose significant challenges. ANNs have been proposed to identify impedance models across varying operating points [15], demonstrating their ability to predict impedance with high accuracy. However, ANNs are heavily dependent on extensive datasets, which makes them less suitable for online impedance identification where data availability is typically limited. Additionally, ANN models are prone to overfitting when data is scarce [16], which can compromise the stability and accuracy of the system analysis. To alleviate the data requirements of standard ANNs, a physics-informed neural network approach was introduced [17], [18], utilizing the physical properties of voltage source converter to compress\\
the network structure and significantly reduce the training load. However, this approach requires explicit expressions of the relationship between operating points and impedance. This is particularly challenging in single-phase vehicle-grid systems due to the complex internal dynamics of the traction converter and its nonlinear relationship with operating points. In summary, while nonlinear black-box impedance identification methods offer significant potential in capturing system dynamics, they face practical challenges in data availability and modeling complexity. Further research is needed to refine these methods and tailor them to the intricate dynamics of railway vehicle-grid systems.

To analyze the stability of the railway vehicle-grid system, selecting the appropriate stability analysis methods based on impedance results is crucial. In the current literature, stability assessments are typically performed by examining the eigenvalue trajectories of the system return ratio matrix, either for a single vehicle [6] or for multiple parallel vehicles [19]. This approach helps to assess system stability. However, these studies usually assume that all vehicles are connected in parallel at a single node, which neglects the impact of line or rail impedance between different vehicles. To address this limitation, the stability of multi-parallel inverter systems has been analyzed to consider the effects of various inverter locations, line parameters, and topological factors on system stability [20]. Nevertheless, the dynamic mobility of vehicles and the frequent changes in their operating conditions mean that analysis methods developed for static multi-parallel inverter systems cannot be directly applied to railway vehicle-grid systems. In addition to these challenges, some studies have explored the stability of railway vehicle-grid systems under single-way double-track power feeding modes [21], while others have investigated stability issues with fully parallel autotransformer traction power supply system [22]. While these studies provide valuable insights into stability under different power supply topologies, the stability of the railway vehicle-grid system under double-sided fully parallel direct power feeding remains unexamined in the existing literature.

To address the issues outlined, this paper proposes a new impedance identification method using a ResFNN network structure combined with SHAP to train a vehicle impedance model, followed by transfer learning to acquire impedance models for other vehicles. Additionally, a CCM-based stability analysis method is introduced to evaluate the stability of multivehicle railway systems under all parallel double-sided feeding mode. The primary innovations of this paper are as follows:

\begin{enumerate}
  \item ResFNN Network with SHAP: The proposed ResFNN, combined with SHAP, is designed specifically for singlephase traction converters. This approach significantly reduces the training data requirements compared to standard FNNs while preserving model accuracy. Furthermore, by leveraging SHAP, this network enhances model interpretability.
  \item CCM-based Stability Analysis Method: The proposed CCM-based stability analysis method incorporates the dynamic mobility of vehicles and their positional distribution. This approach allows for accurate assessment of both low-frequency and high-frequency instability issues\\
within the all parallel double-sided feeding mode.
  \item Integrated Analysis Framework: By integrating the ResFNN network with the stability analysis method, the paper establishes a comprehensive online framework for analyzing the stability of railway vehicle-grid systems. This framework provides the dynamic nature of railway operations for monitoring system stability.\\
The rest of this paper is organized as follows. The proposed ResFNN Network with SHAP for the impedance identification method for vehicles are presented in Section II. Section III details the proposed stability analysis method for evaluating railway vehicle-grid system stability under the double-sided all-parallel power supply mode. Section IV analyzes case studies based on the rail schedule diagram. Section V lists the conclusion.
\end{enumerate}

\section*{II. Impedance Identification of Vehicle}
In this section, we propose a neural network framework to identify vehicle impedance models. Given the challenges of obtaining large-scale impedance datasets in real-world scenarios, we introduce a ResFNN network framework. This framework utilizes SHAP to identify the contributions of different output parameters to impedance, enabling the selection of operating point steps. Subsequently, we leverage transfer learning by using the trained network results as a pretrained model to identify the impedance characteristics of other vehicles. This method is intended to train the model using a small sample dataset while still ensuring the accuracy of the model.

\section*{A. Traction power supply system of vehicle}
Figure 2 shows the topology and control structure of the vehicle power supply system. This system comprises an onboard transformer, line-side rectifier, dc-side filter, motor-side inverter, and traction motor. The line-side rectifier uses a DQ decoupled current control strategy for independent active and reactive power control. The control strategy includes SOGIPLL, dc voltage control, and decoupled DQ-current control. As the vehicle operates, variations in operating point, input voltage $e_{\text{in }}$, and power $P$ affect the vehicle's impedance characteristics. Consequently, we investigated methods to identify these impedance characteristics.

\section*{B. ResFNN architectures}
The flowchart in Fig. 3 outlines the process for training a ResFNN to identify the DQ impedance of multiple vehicles, encompassing a sequence of stages focused on data collection, training, and iterative refinement. The process contains three stages.

\begin{enumerate}
  \item SHAP: At stage 1, by utilizing SHAP in conjunction with a ResFNN, we can quantify the impact of individual features in identifying the DQ impedance model of converters, ultimately aiding in model refinement and understanding.
\end{enumerate}

In the field of machine learning interpretability, SHAP is a popular framework that enables model-agnostic explanations of feature importance [23]-[25]. The methodology is based

\begin{figure}[H]
\begin{center}
  \includegraphics[alt={},max width=\columnwidth]{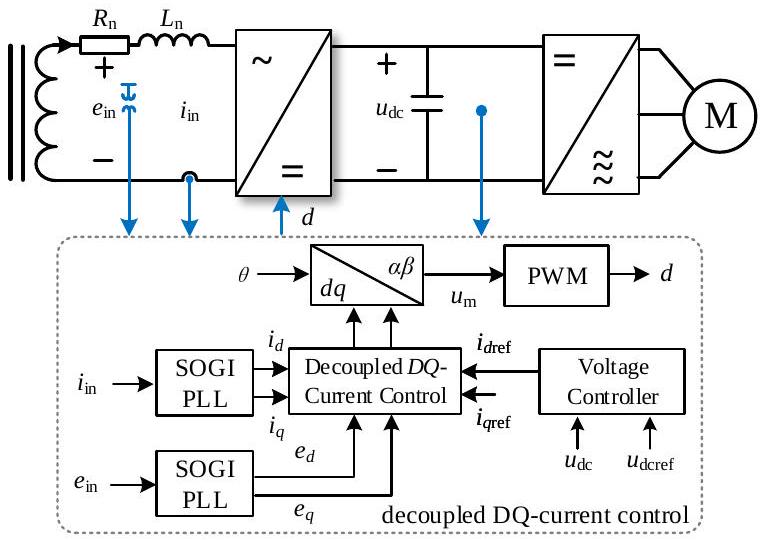}
\caption{Schematic of control dynamics in vehicle traction power supply system.}
\end{center}
\end{figure}

\begin{figure}[H]
\begin{center}
  \includegraphics[alt={},max width=\columnwidth]{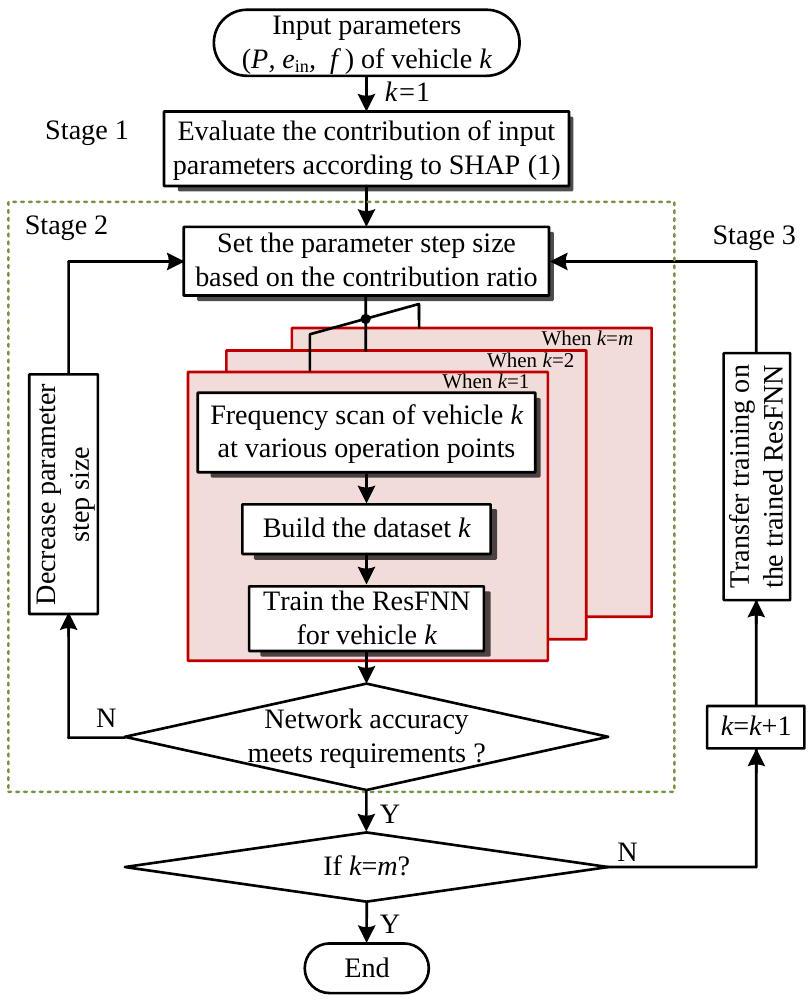}
\caption{Flowchart for modeling DQ impedance of multiple vehicles using ResFNN with SHAP and transfer learning.}
\end{center}
\end{figure}

on Shapley values, a concept derived from cooperative game theory [26], which fairly assigns the contribution of each feature in a predictive model to the overall output. The SHAP framework adheres to the principles of local and global interpretability. Locally, SHAP values explain the contribution of each feature for a specific prediction, while globally, these values aggregate to reveal overall feature importance.

The Shapley value $\phi_{i}$ for a feature $i$ is calculated using the following formula:

\begin{equation*}
\phi_{i}=\sum_{S \subseteq N \backslash\{i\}} \frac{|S|!(|N|-|S|-1)!}{|N|!}[f(S \cup\{i\})-f(S)] \tag{1}
\end{equation*}

\begin{figure}[H]
\begin{center}
  \includegraphics[alt={},max width=\columnwidth]{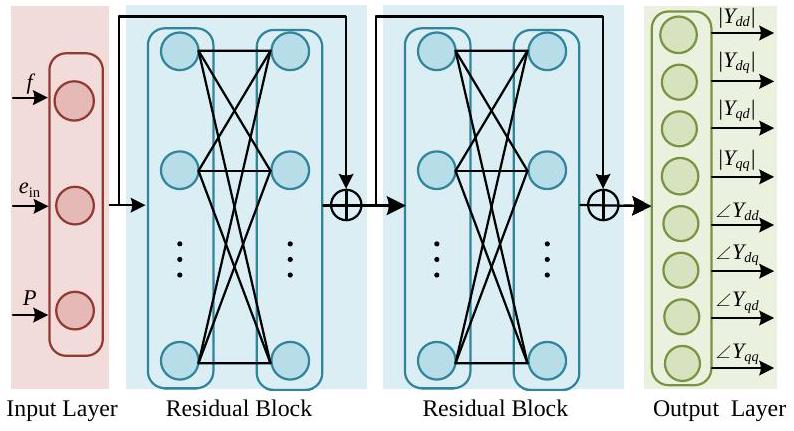}
\caption{Network stucture of ResFNN.}
\end{center}
\end{figure}

where: $N$ is the set of all features, $S$ is a subset of $N$ not containing feature $i,|S|$ is the cardinality of the subset $S$, $f(S)$ represents the predictive model's output when only the features in subset $S$ are used, $f(S \cup\{i\})$ is the model's output when feature $i$ is added to the subset.\\
2) ResFNN: The core of stage 2 involves conducting frequency scans at various operational points for each vehicle, creating a dataset from this collected data, and subsequently training the ResFNN to fit this data. This dataset construction is critical as it ensures the model is trained on relevant and operation-specific data. Following the training, the model's accuracy is assessed against predefined requirements. If the model meets these requirements, the process advances to the next vehicle; otherwise, it may necessitate a return to earlier steps for further data refinement or additional training.

The Residual feedforward neural network used for identifying the DQ impedance model of converters is structured with six layers. It begins with an input layer that receives three parameters: frequency $(f)$, voltage $\left(e_{\text{in }}\right)$, and power $(P)$. Frequency determines the impedance across different points, while voltage and power define the operating state of the vehicle. This is followed by two residual blocks, each containing two hidden layers. These hidden layers, with 64 neurons each, feature skip connections that allow information to bypass intermediate layers, ensuring gradients flow smoothly and training remains efficient. Finally, an output layer provides eight quantities that represent the magnitude and phase values of the DQ impedance for both D and Q components.

The network is trained using MSE as the loss function, which measures the average squared differences between predicted and actual values, ensuring that the training minimizes prediction discrepancies. This function is defined as:

\begin{equation*}
M S E=\frac{1}{n} \sum_{i=1}^{n}\left(\hat{y}_{i}-y_{i}\right)^{2} \tag{2}
\end{equation*}

where $n$ denotes the number of training samples. Optimizers such as Adam dynamically adjust learning rates to minimize this error, ensuring convergence. The ResFNN architecture is particularly well-suited for this application because the residual blocks help preserve important information from earlier layers, making the learning process more efficient and counteracting the vanishing gradient problem. By combining ResFNN with SHAP, the model can offer accurate DQ\\
impedance identification while providing interpretability by revealing feature importance, leading to deeper insights into converter characteristics.\\
3) Transfer learning: The stage 3 allows for the transfer of learned parameters or structures from previously trained models to new models, enhancing efficiency and leveraging shared characteristics among vehicles.

Transfer learning with a pre-trained ResFNN involves leveraging existing knowledge from a previously trained model to quickly adapt and identify the DQ impedance model for other vehicles. The process begins with selecting the ResFNN model already trained on one vehicle's DQ impedance dataset. This pre-trained model contains essential learned features that can be transferred to new vehicles with similar underlying impedance characteristics. The model is then adjusted to fit the specific requirements of the new vehicle's dataset by modifying or fine-tuning its architecture, such as changing the number of neurons or layers in the residual blocks. During training on the new dataset, foundational layers near the input remain frozen to retain critical feature extraction, while the upper layers near the output are fine-tuned to adapt the model's predictive capacity to the new vehicle's characteristics.

This method reduces the overall training time because much of the pre-trained model's structure and weights are reused, requiring fewer computational resources for adaptation. Furthermore, it enhances accuracy because the model quickly converges using established impedance modeling features. This improves overall performance, as the network can adapt its predictions to new vehicles while retaining foundational knowledge. Finally, by preventing overfitting to a single vehicle's dataset, transfer learning enables the model to generalize effectively across different vehicles, providing accurate DQ impedance identification for a broader range of applications.

\section*{C. Training results analysis}
To validate the effectiveness of the proposed ResFNN method, tests are conducted on the traction drive system illustrated in Fig. 2. The dataset for training the network can be obtained using impedance sweep methods or by calculating it through an already constructed impedance model [27]. The electrical system and control parameters of the traction drive system are provided in [22].

The input parameters for the network shown in Fig. 4 include frequency, vehicle input voltage, and operational power. The frequency is defined within the range of $[1,1000] \mathrm{Hz}$. The operational values for the vehicle's input voltage and power are defined within the following ranges:

\[
\left\{\begin{array}{l}
e_{\text{in }} \in[0.9,1.1] \text{ p.u. }  \tag{3}\\
P \in[-1,1] \text{ p.u. }
\end{array}\right.
\]

According to the flowchart illustrated in Fig. 3, the SHAP values are used to determine the contribution of different input features to vehicle impedance. Figure 5 presents the SHAP values for each input feature. The results indicate that frequency $f$ has the most significant average impact on vehicle admittance ( 0.951 ), followed by power ( 0.028 ) and voltage (0.003). Therefore, the construction of the dataset should

\begin{figure}[H]
\begin{center}
  \includegraphics[alt={},max width=\columnwidth]{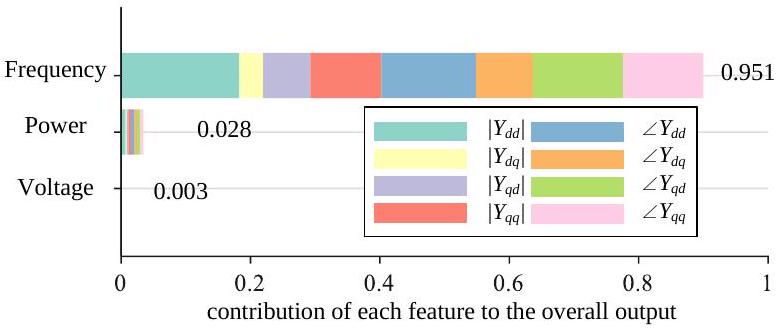}
\caption{Input feature contribution on vehicle impedance.}
\end{center}
\end{figure}

\begin{figure}[H]
\begin{center}
  \includegraphics[alt={},max width=\columnwidth]{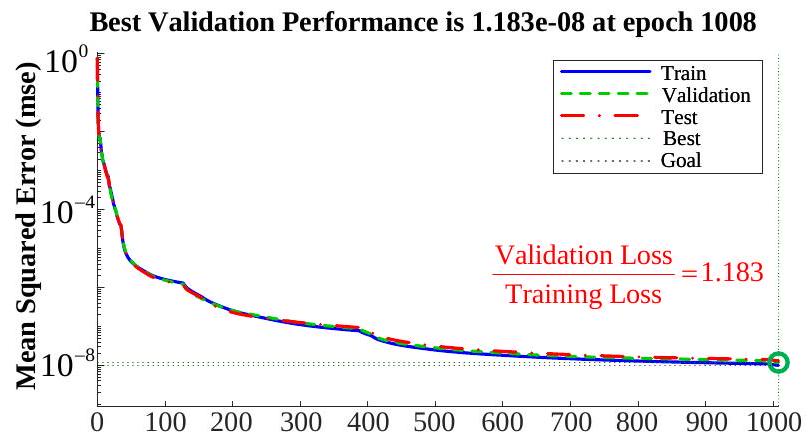}
\caption{ResFNN training process.}
\end{center}
\end{figure}

particularly emphasize the step sizes for frequency and power. It is recommended to decrease the step sizes for frequency and power, while increasing the step size for voltage. Additionally, since the contribution of power is approximately nine times that of voltage, the power step size should be approximately set to one-ninth of the voltage step size. This approach aims to capture more vehicle impedance data at different power points within the dataset. This method is designed to increase the step size of parameters with lower contributions and decrease the step size of parameters with higher contributions, thereby minimizing the dataset size.

After training the model with this dataset, Fig. 6 illustrates the training process of a ResFNN over 1000 epochs, demonstrating how the MSE for the training, validation, and test datasets evolves. As training progresses, all three error curves-training, validation, and test-show a consistent decrease in MSE, suggesting that the model is effectively learning and generalizing well to unseen data. This is further evidenced by the close tracking of the validation and test errors with the training error, indicating that the model is not overfitting, a common concern in neural network training. Besides, the best validation performance, as noted in the figure, is achieved at epoch 1008, with an MSE of approximately $1.183 \times 10^{-8}$. The fact that the training, validation, and test errors approach this goal line towards the end of the training period suggests that the model's performance is aligning well with the predefined objectives. As shown in Fig. 7, after the training, the admittance identification results of vehicle are provided.

In Fig. 8, the performance of three neural network mod-els-FNN, ResFNN, and transfer learned ResFNN-is compared across the MSE of admittance magnitude and phase. The networks are trained using varying dataset sizes to assess\\
\includegraphics[max width=\columnwidth, alt={}, center]{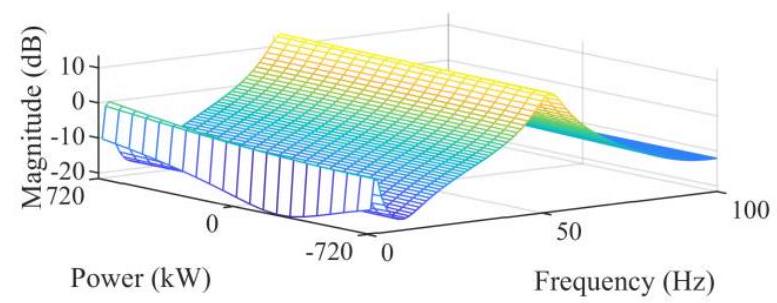}

\begin{figure}[H]
\begin{center}
  \includegraphics[alt={},max width=\columnwidth]{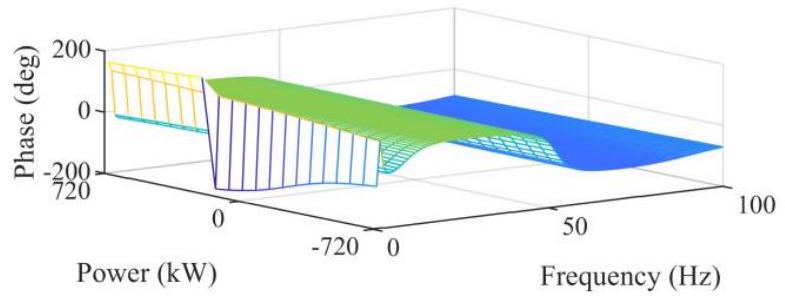}
\caption{Admittance identification results of $Y_{d d}$.}
\end{center}
\end{figure}

\begin{figure}[H]
\begin{center}
  \includegraphics[alt={},max width=\columnwidth]{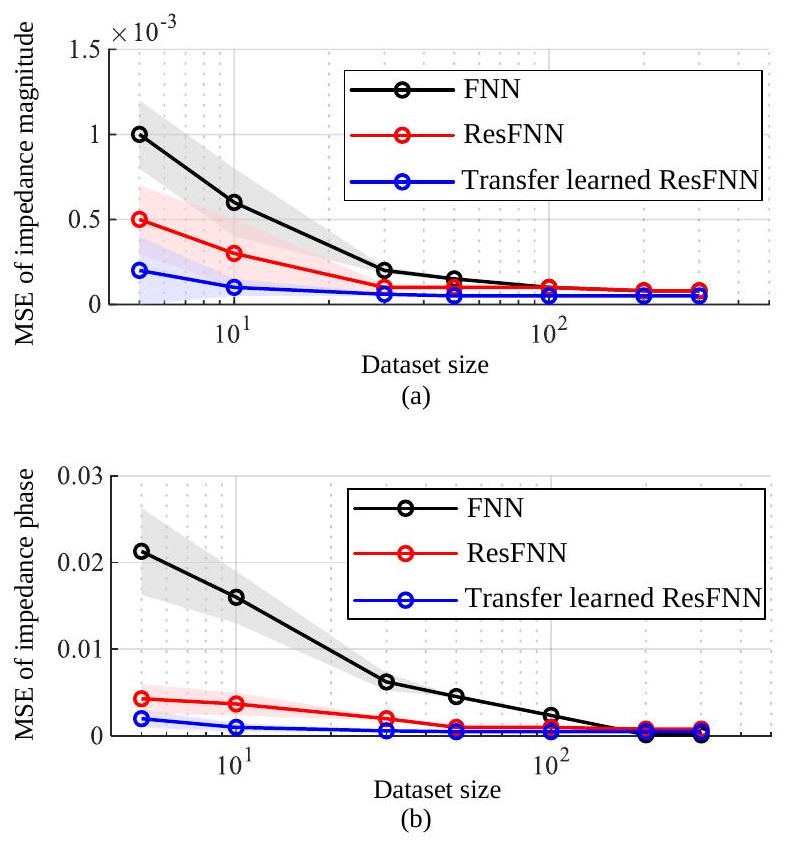}
\caption{Performance evaluations of FNN, ResFNN and transfer learned ResFNN model. (a) MSE of admittance magnitude. (b) MSE of admittance phase.}
\end{center}
\end{figure}

their performance across different data samples. To mitigate the influence of network initialization settings on the training outcomes, each model underwent a training process 10 times, employing distinct random seeds for each iteration. Evaluation results are systematically recorded at the end of each training session. This approach ensures a more robust understanding of each model's capabilities and stability across diverse training conditions.

The performance is plotted against different dataset sizes on a logarithmic scale from about 10 to 100 . The transfer learned ResFNN consistently outperforms the other two models across both metrics and all dataset sizes, indicating a robust advantage from the pre-training which allows it to start with lower errors. This advantage is particularly evident as the dataset size\\
increases, where the transfer learned ResFNN's error remains flat, suggesting an effective use of prior knowledge that needs less data to fine-tune to high accuracy.

The ResFNN also performs better than the basic FNN, especially evident in the phase predictions, where its error decreases steadily and remains below that of the FNN. This trend demonstrates the benefits of residual connections, which help the network learn more complex patterns and improve generalization over the basic FNN architecture. Conversely, the FNN shows the highest error rates but demonstrates improvement as more data is provided. Overall, the analysis of these models in predicting admittance characteristics suggests that transfer learning and residual architectures offer improvements in predictive accuracy, particularly in applications where data is scarce present.

\section*{D. Comparison amongst proposed methods}
Table I presents a comprehensive comparison of the proposed ResFNN with SHAP technique against existing state-of-the-art model identification methods for power converters. This table, along with its detailed description, serves as a practical guide for selecting the most suitable system identification method based on the specific requirements and constraints of the application. The selection criteria should be based on the desired model accuracy, interpretability, operational flexibility, and the characteristics of the available data.

The structural complexity of the models varies, from simple linear combinations in Polytopic models to more complex, layered neural network architectures. The choice of model structure influences the flexibility and applicability of the method across various system identification scenarios. Concerning linear model independence, most methods, with the exceptions of Vector Fitting and Iterative Least Squares, demonstrate independence from linear models. This independence is crucial for adeptly managing the dynamics of nonlinear systems, particularly beneficial in scenarios involving complex, nonlinear behaviors where traditional linear models are inadequate.

In terms of prior knowledge independence, while neural networks (FNN, ANN, RNN) and related models like ResFNN operate independently of prior knowledge and learn directly from data, physics-informed networks and more traditional parametric methods like Vector Fitting necessitate some level of prior knowledge for effective model specification.

Furthermore, in regard to interpretability, Physics-informed Neural Networks and ResFNN with SHAP enhance the interpretability of the models, an essential attribute for applications requiring insight into the decision-making processes of the model. This is in contrast to other neural network approaches, which generally offer lower interpretability.

\section*{III. Stability Analysis Modeling and Framework for Railway Systems}
In this section, we propose a CCM-based method to build the railway system model under double-sided power supply mode. Then, by integrating the ResFNN network with the stability analysis method, the paper establishes a comprehensive online framework for analyzing the stability of railway vehicle-grid systems.

\begin{figure}[H]
\begin{center}
  \includegraphics[alt={},max width=\columnwidth]{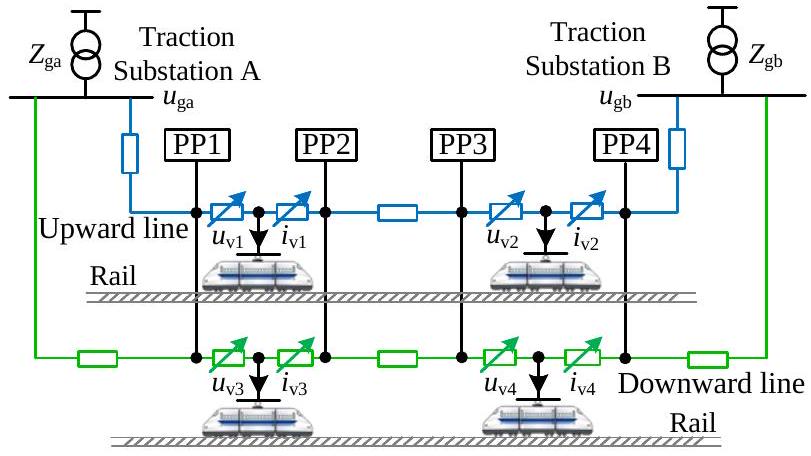}
\caption{Schematic diagram of the railway vehicle-grid system under doublesided all-parallel direct power supply.}
\end{center}
\end{figure}

\begin{figure}[H]
\begin{center}
  \includegraphics[alt={},max width=\columnwidth]{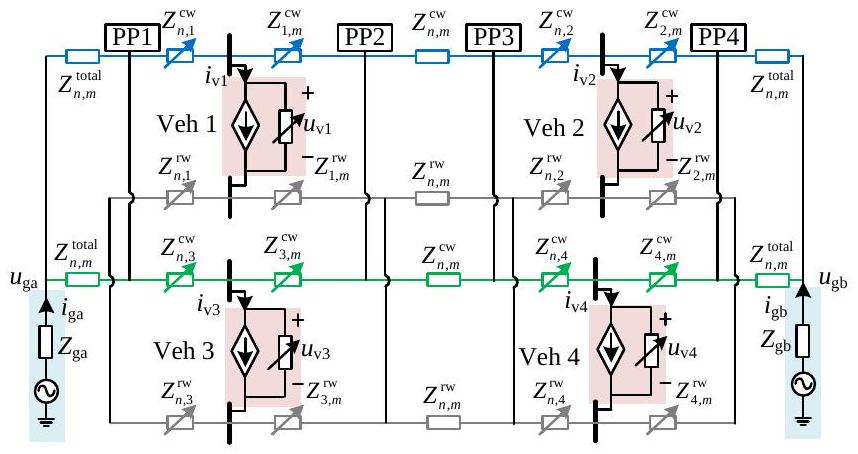}
\caption{Equivalent circuit model of the railway vehicle-grid system.}
\end{center}
\end{figure}

\section*{A. Structure of double-sided all-parallel direct power supply railway system}
Figure 9 illustrates the main components and structure of the vehicle-grid system operating under the double-sided allparallel direct power supply mode. The system comprises two tracks, namely the upward and downward lines, with vehicles simultaneously operational on both. Vehicle power, operational conditions, and locations dynamically change during transit. Vehicles typically operate in traction and braking modes. Furthermore, the all-parallel connection mode enhances system efficiency by establishing equipotentiality between the upward and downward lines through parallel connection points spaced along the tracks. This configuration significantly lowers the traction network's equivalent impedance and increases the reach of the power supply arm.

\section*{B. All-parallel double-sided direct power supply system model}
Figure 10 presents the equivalent circuit model of a railway vehicle-grid system. This model illustrates the interaction between four railway vehicles (Veh 1, Veh 2, Veh 3, Veh 4) and the power grid. The system is divided into multiple sections, which are connected by parallel points (PP1, PP2, PP3, PP4) that serve as the connection points between the upward and downward lines.

The subscript 'ga(b)' represents the equivalent voltage on the secondary side of traction substation $\mathrm{A}(\mathrm{B})$, and $Z_{\mathrm{g}}$ denotes the equivalent impedance of the traction substation. The superscript 'cw' indicates the catenary impedance, 'rw' represents

\begin{table*}[!t]
\centering
\caption{Comparison of the proposed ResFNN with the state-of-the-art model identification method of power converters}
\resizebox{\textwidth}{!}{%
\begin{tabular}{|l|l|l|l|l|l|l|}
\hline
Method & Model Structure & Linear Model Independent & Prior Knowledge Independent & Transfer Function Expression & Interpretability & Identification Object \\
\hline
Polytopic [12], [13] & Multiple linear models combined & Yes & Partially & No & N/A & dc-dc converter \\
\hline
Hammerstein [28] & Combination of a nonlinear function and linear models & Partially (depends on linear dynamic part) & Partially & No for nonlinear part, Yes for linear part & separate of nonlinear and linear feature & dc-dc converter \\
\hline
Vector Fitting [29], [30] & Transfer function approximations & No & No & Yes & poles and zeros representation & multiple inverter system \\
\hline
NARX [10] & Nonlinear autoregressive with exogenous inputs & Yes & Partially & No & embed a prior knowledge & dc-dc converter \\
\hline
Iterative Least Squares [9] & Parametric models (linear or nonlinear) & No & No & Yes & reveal parameters sensitivity on model output & half-bridge converter \\
\hline
RNN [14] & Recurrent neural networks & Yes & Yes & No & N/A & three phase converter \\
\hline
Neural Network (ANN [15], FNN [16]) & Layered neural networks & Yes & Yes & No & N/A & three phase converter \\
\hline
Physics-informed Neural Network [17], [18] & Neural networks integrated with physical laws & Yes & No & No & incorporate physical laws into network & dc-dc converter, three phase converter \\
\hline
ResFNN with SHAP & Residual feedforward neural network with interpretability tools & Yes & Yes & No & input feature contribution on model output & single phase converter \\
\hline
\end{tabular}
}
\end{table*}

the rail impedance, and 'total' considers both catenary and rail impedances simultaneously. Since the vehicle is moving along the track, variable impedance is used to model the line impedance at different vehicle positions.

The impedance between two PPs is denoted as $Z_{n, m}$. When a vehicle is situated in the interval between these two PPs, the electrical characteristics of the system can be modeled using specific impedance values. Specifically, the impedance from the left PP to the vehicle node is represented as $Z_{n, k}$, while the impedance from the right PP to the vehicle node is represented as $Z_{k, m}$. These impedance values are not fixed but vary depending on the exact position $l$ of the vehicle within the interval between the PPs. The relationship between these impedances and the vehicle's position can be mathematically expressed as follows:

\begin{gather*}
Z_{n, k}=\frac{l}{L}\left(r_{\mathrm{u}}+j \cdot x_{\mathrm{u}}\right)  \tag{4}\\
Z_{k, m}=\left(1-\frac{l}{L}\right)\left(r_{\mathrm{u}}+j \cdot x_{\mathrm{u}}\right) \tag{5}
\end{gather*}

where $L$ represents the total length of a PP segment, and $z_{\mathrm{u}}=r_{\mathrm{u}}+j x_{\mathrm{u}}$ is the unit length impedance of the equivalent circuit of the traction network. Here, $r_{\mathrm{u}}$ denotes the resistance per unit length, while $x_{\mathrm{u}}$ represents the reactance per unit length. The combined effect of these two components, resistance and reactance, gives the total impedance per unit length, which is crucial for accurately modeling the voltage drop and power loss in the traction network as the vehicle moves along the segment.

According to Fig. 10, the nodes chosen for analysis encompass the locations of vehicles and traction substations, as\\
well as the parallel points. The node voltage is presented by $\boldsymbol{u}=\left[\boldsymbol{u}_{\mathrm{v} 1}, \boldsymbol{u}_{\mathrm{v} 2}, \boldsymbol{u}_{\mathrm{v} 3}, \boldsymbol{u}_{\mathrm{v} 4}, \boldsymbol{u}_{\mathrm{PP} 1}, \boldsymbol{u}_{\mathrm{PP} 2}, \boldsymbol{u}_{\mathrm{PP} 3}, \boldsymbol{u}_{\mathrm{PP} 4}, \boldsymbol{u}_{\mathrm{ga}}, \boldsymbol{u}_{\mathrm{gb}}\right]$, and the node current by $\boldsymbol{i}=\left[\boldsymbol{i}_{\mathrm{v} 1}, \boldsymbol{i}_{\mathrm{v} 2}, \boldsymbol{i}_{\mathrm{v} 3}, \boldsymbol{i}_{\mathrm{v} 4}, \boldsymbol{i}_{\mathrm{PP} 1}, \boldsymbol{i}_{\mathrm{PP} 2}\right.$, $\left.\boldsymbol{i}_{\mathrm{PP} 3}, \boldsymbol{i}_{\mathrm{PP} 4}, \boldsymbol{i}_{\mathrm{ga}}, \boldsymbol{i}_{\mathrm{gb}}\right]$.

The railway vehicle-grid system can be divided into the vehicle subsystem and the traction network subsystem for analyzing. The vehicle subsystem includes all vehicles, and the traction network subsystem consists of parallel points and the traction substation. The node voltage of vehicle subsystem is represented as $\boldsymbol{u}_{\mathrm{vs}}=\left[u_{\mathrm{v} 1}, u_{\mathrm{v} 2}, u_{\mathrm{v} 3}, u_{\mathrm{v} 4}\right]^{\mathrm{T}}$, and node current as $\boldsymbol{i}_{\mathrm{vs}}=\left[i_{\mathrm{v} 1}, i_{\mathrm{v} 2}, i_{\mathrm{v} 3}, i_{\mathrm{v} 4}\right]^{\mathrm{T}}$. The node voltage of traction subsystem is represented as $\boldsymbol{u}_{\mathrm{ts}}=\left[u_{\mathrm{PP} 1}, u_{\mathrm{PP} 2}, u_{\mathrm{PP} 3}, u_{\mathrm{PP} 4}\right.$, $\left.u_{\mathrm{ga}}, u_{\mathrm{gb}}\right]^{\mathrm{T}}$, and the corresponding currents are $\boldsymbol{i}_{\mathrm{ts}}=\left[i_{\mathrm{PP} 1}\right.$, $\left.i_{\text{PP2 }}, i_{\text{PP3 }}, i_{\text{PP4 }}, i_{\text{ga }}, i_{\text{gb }}\right]^{\mathrm{T}}$. Based on the defined nodes, the node admittance matrix $\boldsymbol{Y}_{\text{sys }}$ of the vehicle-grid system can be constructed.

\[
\underbrace{\left[\begin{array}{c}
\boldsymbol{i}_{\mathrm{vs}}  \tag{6}\\
\boldsymbol{i}_{\mathrm{ts}}
\end{array}\right]}_{\boldsymbol{i}}=\boldsymbol{Y}_{\text{sys }} \underbrace{\left[\begin{array}{c}
\boldsymbol{u}_{\mathrm{vs}} \\
\boldsymbol{u}_{\mathrm{ts}}
\end{array}\right]}_{\boldsymbol{u}}
\]

where the expression for $\boldsymbol{Y}_{\text{sys }}$ is provided in the appendix.\\
Then, the node admittance matrix $\boldsymbol{Y}_{\text{sys }}$ is adopted to eliminate nodes without connected vehicles using the Kron reduction [31]. Then, (6) can be expressed as

\[
\left[\begin{array}{c}
\boldsymbol{i}_{\mathrm{vs}}  \tag{7}\\
\boldsymbol{i}_{\mathrm{ts}}
\end{array}\right]=\left[\begin{array}{ll}
\boldsymbol{Y}_{\mathrm{mm}} & \boldsymbol{Y}_{\mathrm{mn}} \\
\boldsymbol{Y}_{\mathrm{nm}} & \boldsymbol{Y}_{\mathrm{nn}}
\end{array}\right]\left[\begin{array}{l}
\boldsymbol{u}_{\mathrm{vs}} \\
\boldsymbol{u}_{\mathrm{ts}}
\end{array}\right]
\]

Eliminating $\boldsymbol{u}_{\mathrm{ts}}$ and $\boldsymbol{i}_{\mathrm{ts}}$ from the above matrix, the admittance matrix between $\boldsymbol{u}_{\mathrm{vs}}$ and $\boldsymbol{i}_{\mathrm{vs}}$ can be derived

\begin{equation*}
\boldsymbol{i}_{\mathrm{vs}}=\underbrace{\left(\boldsymbol{Y}_{\mathrm{mm}}-\boldsymbol{Y}_{\mathrm{mn}} \boldsymbol{Y}_{\mathrm{nn}}^{-1} \boldsymbol{Y}_{\mathrm{nm}}\right)}_{\boldsymbol{Y}_{\mathrm{ts}}} \boldsymbol{u}_{\mathrm{vs}} \tag{8}
\end{equation*}

where $\boldsymbol{Y}_{\text{ts }}$ represents the equivalent traction network subsystem model constructed from the selected vehicle nodes. According to the equivalent circuit of the vehicle network system shown in Fig. 10, based on the selected node voltages, the node admittance matrix of the vehicle network system is provided in Appendix.

\section*{C. Vehicle subsystem model}
Figure 11 illustrates the electrical connection of the railway vehicle-grid system. The vehicle subsystem, represented by $\boldsymbol{Y}_{\text{vsdq }}$, comprises multiple vehicles, each denoted as Vehicle 1 through Vehicle n. Each vehicle has its own internal admittance $\boldsymbol{Y}_{\mathrm{v} 1 d q}$ to $\boldsymbol{Y}_{\mathrm{v} n d q}$, which are connected to the traction network subsystem through designated ports. The steps to build the vehicle subsystem are provided below.

\begin{enumerate}
  \item Step 1. Power flow calculation for traction power systems: This process begins with collecting detailed train schedule information, which includes the departure and arrival times, intermediate stops, and expected travel times between these stops for each train. Using this schedule data, the exact position of each train within the network at any given time can be calculated by mapping the progress of each train along its route. With this information, a load flow calculation program for the traction power supply system is executed to obtain the line-side voltage and phase for each vehicle [32]. These voltage and phase values are critical for subsequent analyses and ensure that the electrical characteristics of the system are accurately represented for further evaluation.
  \item Step 2. Obtain vehicle admittance data using the trained ResFNN network model: Admittance data for each vehicle can be obtained by inputting the line-side voltages, derived from system power flow calculations, and the operating power of the vehicles into the ResFNN model trained in Section II.
  \item Step 3. Transform DQ impedance of vehicle from local to global frame: When modeling the DQ impedance of multiple vehicles, each vehicle's impedance is initially defined in its own $d q$ frame. To analyze these models together, they need to be transformed into a global $d q$ frame, typically a reference synchronous frame based on the grid frequency. Each vehicle's DQ impedance model, represented as a matrix $Y_{d q, i}(s)$, is transformed using a coordinate transformation. Assuming there is a phase angle $\theta_{i}$ between the local $d q$ frame of the $i$-th vehicle and the global reference frame, the transformation matrix $T_{i}\left(\theta_{i}\right)$ is used [33]. The impedance matrix in the local $d q$ frame is transformed into the global reference $d q$ frame using $\boldsymbol{Y}_{\mathrm{v} i d q}^{\mathrm{g}}=T_{d q}^{-1} \boldsymbol{Y}_{\mathrm{v} i d q}^{l} T_{d q}$, where $T_{d q}^{-1}$ is the inverse of $T_{d q}$. This process is repeated for each vehicle, converting all their impedance models to the global reference $d q$ frame. Once transformed, these models can be combined for further analysis.
  \item Step 4. Build the admittance matrix of vehicle subsystem: Once the DQ impedance models for each vehicle are transformed into the global reference $d q$ frame, the next step involves integrating these individual impedance models into a subsystem model. This is achieved by placing each inverter's
\end{enumerate}

\begin{figure}[H]
\begin{center}
  \includegraphics[alt={},max width=\columnwidth]{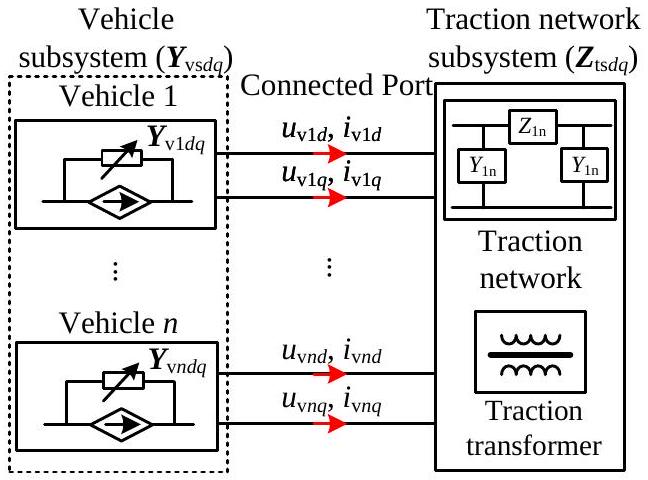}
\caption{Electrical connection of the railway vehicle-grid system.}
\end{center}
\end{figure}

impedance model onto a block-diagonal matrix, representing the entire vehicle subsystem.

\[
\underbrace{\left[\begin{array}{c}
\boldsymbol{i}_{\mathrm{v} 1 d q}  \tag{9}\\
\boldsymbol{i}_{\mathrm{v} 2 d q} \\
\vdots \\
\boldsymbol{i}_{\mathrm{v} n d q}
\end{array}\right]}_{\boldsymbol{i}_{\mathrm{vs} d q}}=\underbrace{\left[\begin{array}{cccc}
\boldsymbol{Y}_{\mathrm{v} 1 d q}^{\mathrm{g}} & & & \\
& \boldsymbol{Y}_{\mathrm{v} 2 d q}^{\mathrm{g}} & & \\
& & \ddots & \\
& & & \boldsymbol{Y}_{\mathrm{v} n d q}^{\mathrm{g}}
\end{array}\right]}_{\boldsymbol{Y}_{\mathrm{vs} d q}} \underbrace{\left[\begin{array}{c}
\boldsymbol{u}_{\mathrm{v} 1 d q} \\
\boldsymbol{u}_{\mathrm{v} 2 d q} \\
\vdots \\
\boldsymbol{u}_{\mathrm{v} n d q}
\end{array}\right]}_{\boldsymbol{u}_{\mathrm{vs} d q}}
\]

\section*{D. Railway vehicle-grid system stability analysis}
Figure 12 presents a flowchart detailing the stability analysis process for a railway vehicle-grid system. The process begins with the system data collection phase, where parameters of the vehicle-grid system are gathered. The railway train schedule diagram is then divided into multiple time points $m$, and a specific time point $t_{i}$ is selected for detailed analysis.

The system is partitioned by multiple points of connection into two subsystems: Passive and Active. For the passive subsystem, the traction network nodal admittance matrix $\boldsymbol{Y}_{\text{sys }}$ is constructed and then partitioned. Subsequently, the vehicle node matrix $\boldsymbol{Y}_{\mathrm{s}}$ is extracted using Kron reduction. Meanwhile, the active subsystem involves performing a power flow analysis of the traction power supply system to determine power $P$ and voltage $e_{\text{in }}$. Operation points are input into the trained ResFNN to obtain $\boldsymbol{Y}_{i d q}$, aligning the vehicle admittance model with the global frame and constructing the admittance matrix of the vehicle subsystem $\boldsymbol{Y}_{\mathrm{ts} d q}$.

Following these steps, the system return-ratio matrix $\boldsymbol{L}_{\mathrm{rt}}$ is built, and the GNSC stability analysis is conducted. The vehicle grid system return ratio matrix can be represented as:

\begin{equation*}
\boldsymbol{L}_{\mathrm{rt}}=\boldsymbol{Z}_{\mathrm{ts} d q} \boldsymbol{Y}_{\mathrm{vs} d q} \tag{10}
\end{equation*}

The GNSC involves plotting the eigenvalues of the returnratio matrix on the complex plane and evaluating their encirclements around the critical point $(-1,0)[34]$. This criterion is used to determine whether the system will remain stable under various operating conditions. Stability margins are critical indicators of system stability. The gain margin is determined by checking where the phase plot crosses -180 degrees [35]. At this frequency, the difference between the magnitude at

\begin{figure}[H]
\begin{center}
  \includegraphics[alt={},max width=\columnwidth,max height=0.75\textheight,keepaspectratio]{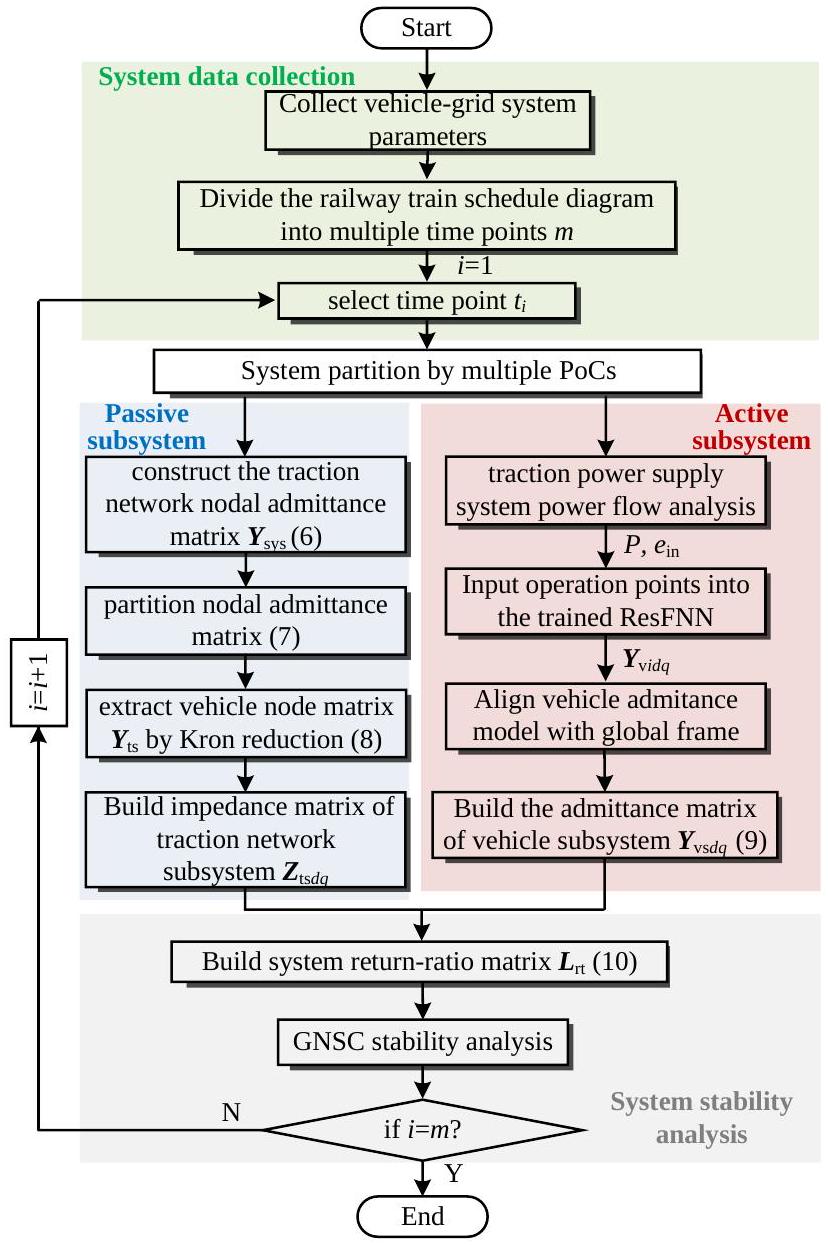}
\caption{Railway vehicle-grid system stability analysis process.}
\end{center}
\end{figure}

this point and 0 dB is the gain margin. A positive gain margin indicates a stable system.

The process iterates by checking if $i=m$. If not, $i$ is incremented, and the analysis is repeated for the next time point $t_{i}$. If $i=m$, the stability analysis concludes, and the process ends. This flowchart provides a structured and systematic approach to analyzing the stability of the railway vehicle-grid system.

\section*{IV. Case Studies}
This section validates the proposed impedance identification and stability analysis method based on the train schedules. Table II provides the railway traction power supply system parameters.

\section*{A. System Description}
Figure 13(a) illustrates the timing and positioning of each train within the network, based on the schedule data that includes departure times, arrival times, and stops at intermediate stations. The diagram shows the movement of multiple vehicles $\left(\mathrm{v}_{1}, \mathrm{v}_{2}, \mathrm{v}_{3}, \mathrm{v}_{4}, \mathrm{v}_{5}, \mathrm{v}_{6}\right)$ between two traction power substations (TPSS 1 and TPSS 2) over time. The horizontal axis represents time in hours, spanning from 9:30 to 10:00,

\begin{table}[H]
\begin{center}
\caption{The Parameters of the Traction Power Supply System}
\begin{tabular}{ccc}
\hline\hline
Item & Parameters & Values \\
\hline
 & Traction network voltage & 27.5 kV \\
 & Traction transformer rated power & 40 MVA \\
Traction power & Overhead contact line & CTMH150 \\
supply system & Power supply arm & 40 km \\
 & Number of all-parallel points & 4 \\
 & Rails & CHN 60 \\
\hline\hline
\end{tabular}
\end{center}
\end{table}

\begin{figure}[H]
\begin{center}
  \includegraphics[alt={},max width=\columnwidth]{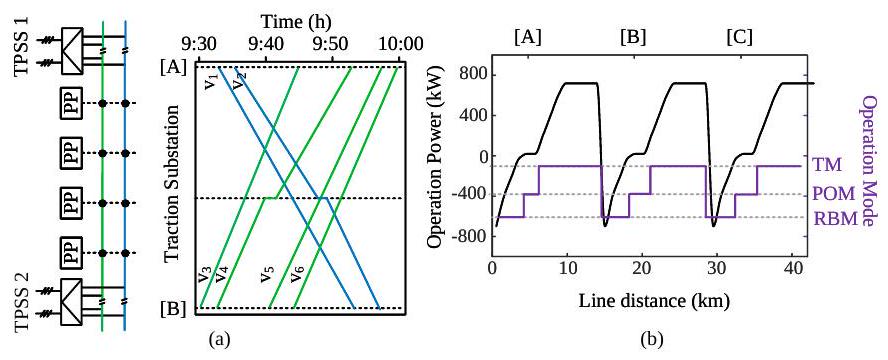}
\caption{Operational characteristics of the railway vehicle-grid system. (a) Vehicle trajectories and positions over time. (b) Operation power and mode along the line distance.}
\end{center}
\end{figure}

while the vertical axis represents the positions of the traction substations $[\mathrm{A}]$ and $[\mathrm{B}]$. The sloped lines indicate the trajectories of the vehicles as they travel between the substations, with green lines representing upward vehicle paths and blue lines representing downward set. Fig. 13(b) depicts the operation power and mode of a vehicle at various line positions. The horizontal axis shows the line distance in kilometers, with intervals marked at every 10 km up to 40 km . The left vertical axis quantifies the operation power in kilowatts, ranging from -800 kW to 800 kW . The right vertical axis indicates the operation mode. The black curve represents the operation power of the vehicle, and the purple line shows the corresponding operation modes, including traction mode (TM), power off mode (POM), and regenerative braking mode (RBM) [36]. Overall, this figure illustrates how train schedules and vehicle operation modes are managed within the railway vehicle-grid system, highlighting the dynamic interaction between train movements and power consumption over time and distance.

To analyze the stability of the vehicle-network system, five time points corresponding to Cases 1 through 5 are shown in Table III. These points are drawn from the timetable in the appendix in Table IV, where vehicle operation power and location details are also provided.

\section*{B. System Stability Analysis}
In Case 1, the railway vehicle-grid system incorporates 4 vehicles, thus vehicle subsystem matrix in the DQ coordinate system is of eight order. According to formulas (12), the return ratio matrix of the vehicle-grid system is also eight order, necessitating the analysis of eight eigenlocus. As depicted in Fig. 14(a), eigenloci $\Lambda_{1}$ exhibits the largest magnitude, indicating its predominant role in system stability. Besides, eigenloci $\Lambda_{1}$ does not intersect the 0 dB line, indicating system stability. It is noted that eigenloci $\Lambda_{1}$ exhibits two amplitude

\begin{table}[H]
\begin{center}
\caption{Case design and stability analysis results}
\begin{tabular}{|l|l|l|l|l|}
\hline
Case & Time & Vehicle at downward line & Vehicle at upward line & Stability analysis results \\
\hline
Case 1 & 9:35 & $\mathrm{v}_{1}, \mathrm{v}_{2}$ & $\mathrm{v}_{3}, \mathrm{v}_{4}$ & stable \\
\hline
Case 2 & 9:41 & $\mathrm{v}_{1}, \mathrm{v}_{2}$ & $\mathrm{v}_{3}, \mathrm{v}_{4}, \mathrm{v}_{5}$ & critical stable \\
\hline
Case 3 & 9:45 & $\mathrm{v}_{1}, \mathrm{v}_{2}$ & $\mathrm{v}_{3}, \mathrm{v}_{4}, \mathrm{v}_{5}$ & unstable at 2.5 Hz \\
\hline
Case 4 & 9:50 & $\mathrm{v}_{1}, \mathrm{v}_{2}$ & $\mathrm{v}_{4}, \mathrm{v}_{5}, \mathrm{v}_{6}$ & unstable at 133 Hz \\
\hline
Case 5 & 9:53 & $\mathrm{v}_{1}, \mathrm{v}_{2}$ & $\mathrm{v}_{5}, \mathrm{v}_{6}$ & stable \\
\hline
\end{tabular}
\end{center}
\end{table}

\begin{figure}[H]
\begin{center}
  \includegraphics[alt={},max width=\columnwidth,max height=0.75\textheight,keepaspectratio]{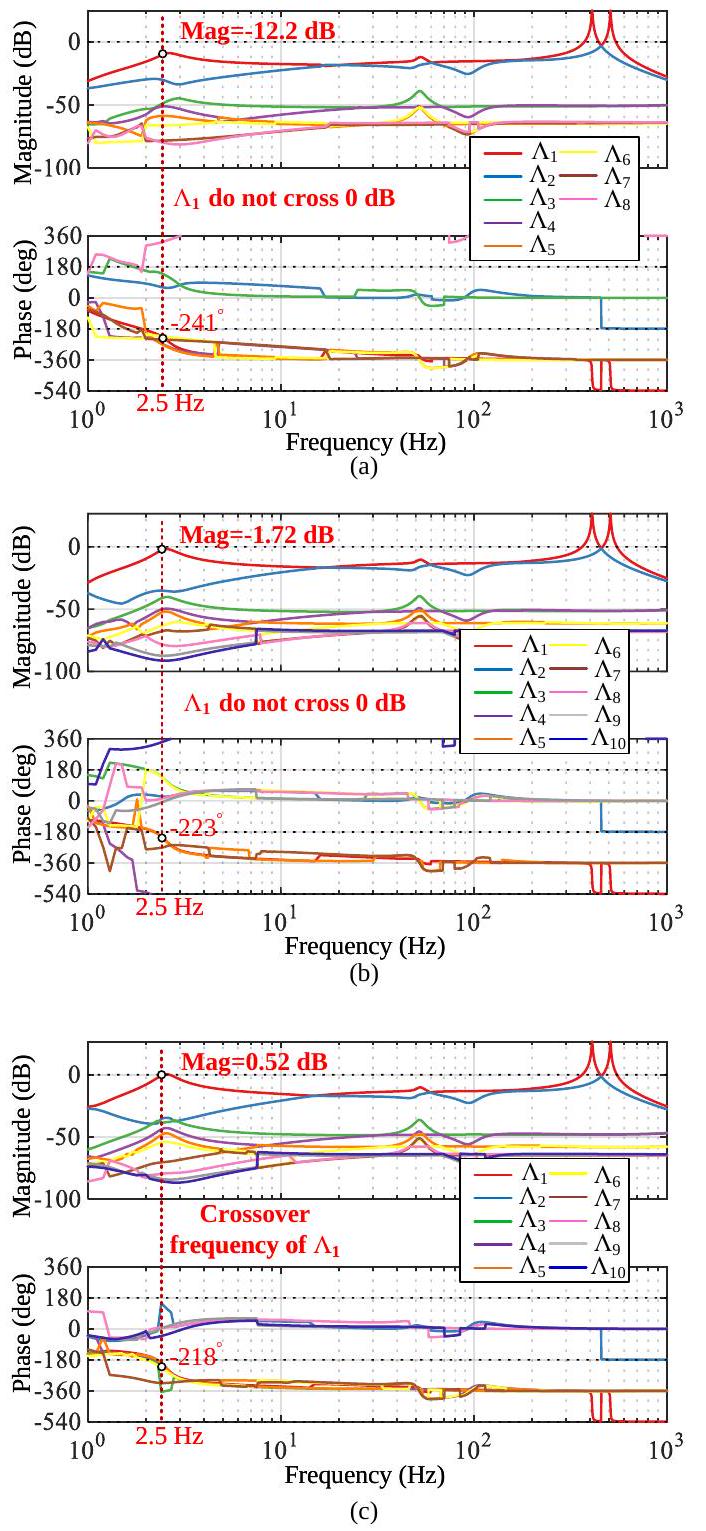}
\caption{Bode diagrams of the eigenloci based on vehicle-grid system model. (a) Case 1. (b) Case 2. (c) Case 3.}
\end{center}
\end{figure}

peaks in the high-frequency range, which are caused by the intrinsic resonance frequencies of the traction network. When the switching harmonics of the vehicle are close to these resonance frequencies, the system may experience harmonic resonance issues.

The stability analysis results for Case 2 are shown in Fig.

\begin{figure}[H]
\begin{center}
  \includegraphics[alt={},max width=\columnwidth,max height=0.75\textheight,keepaspectratio]{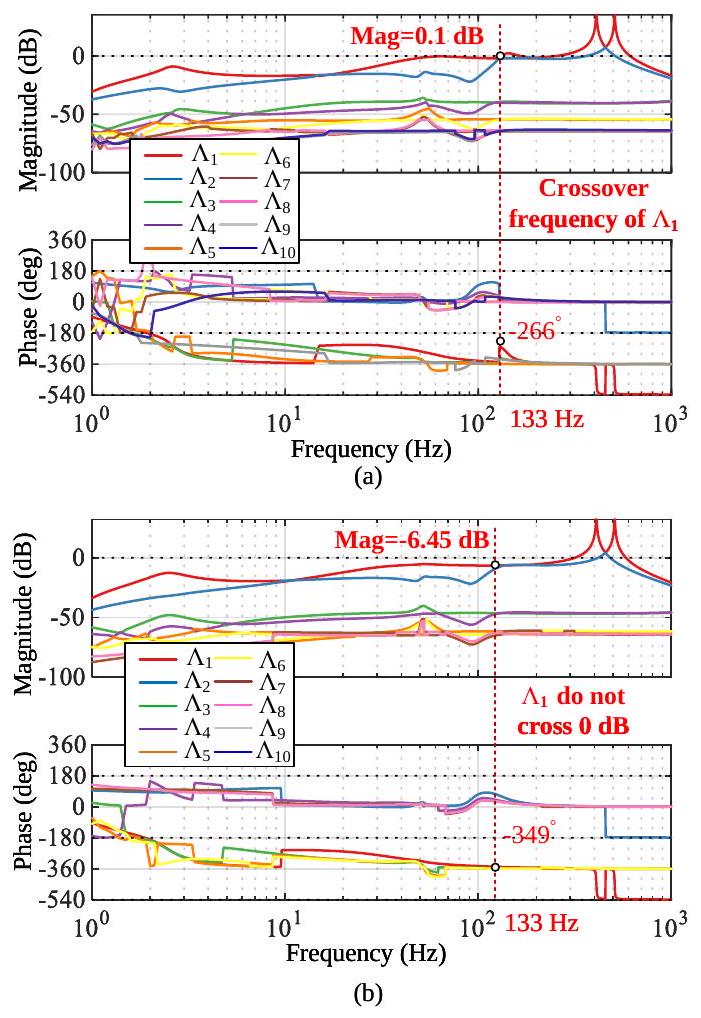}
\caption{Bode diagrams of the eigenloci based on vehicle-grid system model. (a) Case 4. (b) Case 5.}
\end{center}
\end{figure}

14(b). This configuration results in a ten order vehicle subsystem matrix in the DQ coordinate system and a correspondingly ten order return ratio matrix. Therefore, ten eigenlocus are analyzed. In Fig. 14(b), eigenloci $\Lambda_{1}$ has a greater amplitude than the other eigenloci, which similarly indicates its dominant role in the system's stability. Moreover, eigenloci $\Lambda_{1}$ nearly reaches 0 dB at approximately 2.5 Hz , with a peak amplitude of -1.72 dB , remaining below 0 dB . This suggests that the trajectory of eigenloci $\Lambda_{1}$ is very close to the ( $-1, j 0$ ) point without encirclement. Therefore, the system is critically stable, with a low stability margin at Case 2.

Railway vehicle grid system stability analysis results for Case 3 are depicted in Fig. 14(c), where eigenloci $\Lambda_{1}$ crosses the 0 dB threshold with an amplitude of 0.52 dB at 2.5 Hz . Besides, the phase of eigenloci $\Lambda_{1}$ at 2.5 Hz is $-218^{\circ}$, surpassing $-180^{\circ}$. Based on the GNSC stability criterion, the system is unstable.

As illustrated in Fig. 15(a) for Case 4, eigenloci $\Lambda_{1}$ intersects the 0 dB line at 133 Hz with the magnitude of 0.1 dB , and its phase is $-266^{\circ}$. Since the phase exceeding $-180^{\circ}$, according to the GNSC stability criterion, the system is unstable.

The stability analysis results for Case 5 are depicted in Fig. 15(b). In contrast to Case 4, eigenloci $\Lambda_{1}$ at 133 Hz has the magnitude of -6.45 dB and does not intersect the 0 dB line, demonstrating that the system can maintain stability.

\section*{C. Case Validation}
Figure 16(a) displays the validation results for the railway vehicle-grid system in Case 1. By sequentially connecting ve-

\begin{figure}[H]
\begin{center}
  \includegraphics[alt={},max width=\columnwidth,max height=0.75\textheight,keepaspectratio]{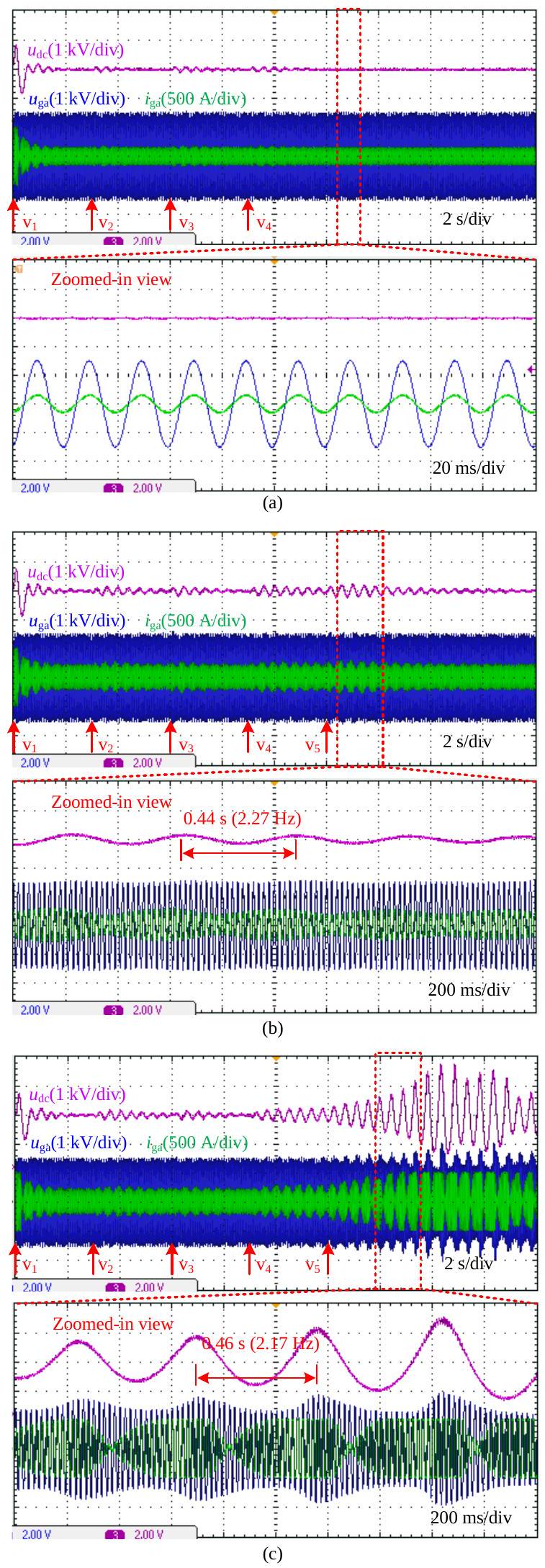}
\caption{Waveforms of line-side voltage and current, and dc-link voltage of vehicle. (a) Case 1. (b) Case 2. (c) Case 3.}
\end{center}
\end{figure}

hicles to the system, the line-side voltage, current waveforms,\\
and the dc-side voltage waveform of the vehicle are observed to evaluate system stability. From Fig. 16(a), it is evident that the system waveforms remain stable as four vehicles are connected sequentially, evidenced by constant DC-side voltages and line-side voltages and currents operating at unity power factor. These results are consistent with those shown in Table III for Case 1.

Figure 16(b) depicts the results for Case 2. The system remained stable with four connected vehicles but exhibited constant amplitude oscillations upon connecting a fifth. The oscillation will persist for several cycles before gradually decaying. This indicates that the system is in a critically stable state at this time. Furthermore, the frequency of oscillation is 2.27 Hz , which aligns with the stability results for Case 2 shown in Table III.

Figure 16(c) displays the results for Case 3, showing that oscillations already occur when four vehicles are sequentially connected. Upon connecting the fifth vehicle, the waveform diverges, accompanied by substantial oscillations in the DCside voltage. At the same time, the line-side voltage and current also exhibit fluctuations without returning to a stable state, indicating that the system is unstable. The waveforms show an oscillation frequency of 2.17 Hz , consistent with the stability results in Table III for Case 3.

Figure 17(a) illustrates the results for Case 4. The system is stable as four vehicles are sequentially connected. However, when the sixth vehicle is connected, the harmonic components in the waveform gradually increase. By analyzing the oscillatory part of the waveform, it is evident that there is a 125 Hz harmonic component in the DC-side voltage. This indicates the presence of harmonic instability in the high-frequency segments of the system. The stability results in Case 4 align with those shown in Table III.

Figure 17(b) shows the results for Case 5. The system is stable as three vehicles are sequentially connected. When a sixth vehicle is connected, the system remains stable without any oscillatory issues. The stability results for Case 5 are consistent with those shown in Table III.

\section*{V. Conclusion}
The paper proposes a ResFNN Network with SHAP for the impedance identification of single-phase traction converters. The proposed technique reduces the training data requirements compared to standard FNNs while maintaining model accuracy. Additionally, a CCM-based method is proposed to analyze the stability of multi-vehicle railway system under the double-sided power supply mode, incorporating the dynamic mobility of vehicles and their positional distribution. The key advantages of the proposed methods are summarized as follows:

\begin{enumerate}
  \item In terms of impedance identification performance, the proposed ResFNN method, enhanced by SHAP for modelagnostic explanations of feature importance, optimizes the input parameters' step size based on their contribution values. This optimization constructs a dataset suitable for small samples. Training results indicate that ResFNN outperforms basic FNN in predictive accuracy, especially in scenarios with limited data.
\end{enumerate}

\begin{figure}[H]
\begin{center}
  \includegraphics[alt={},max width=\columnwidth,max height=0.75\textheight,keepaspectratio]{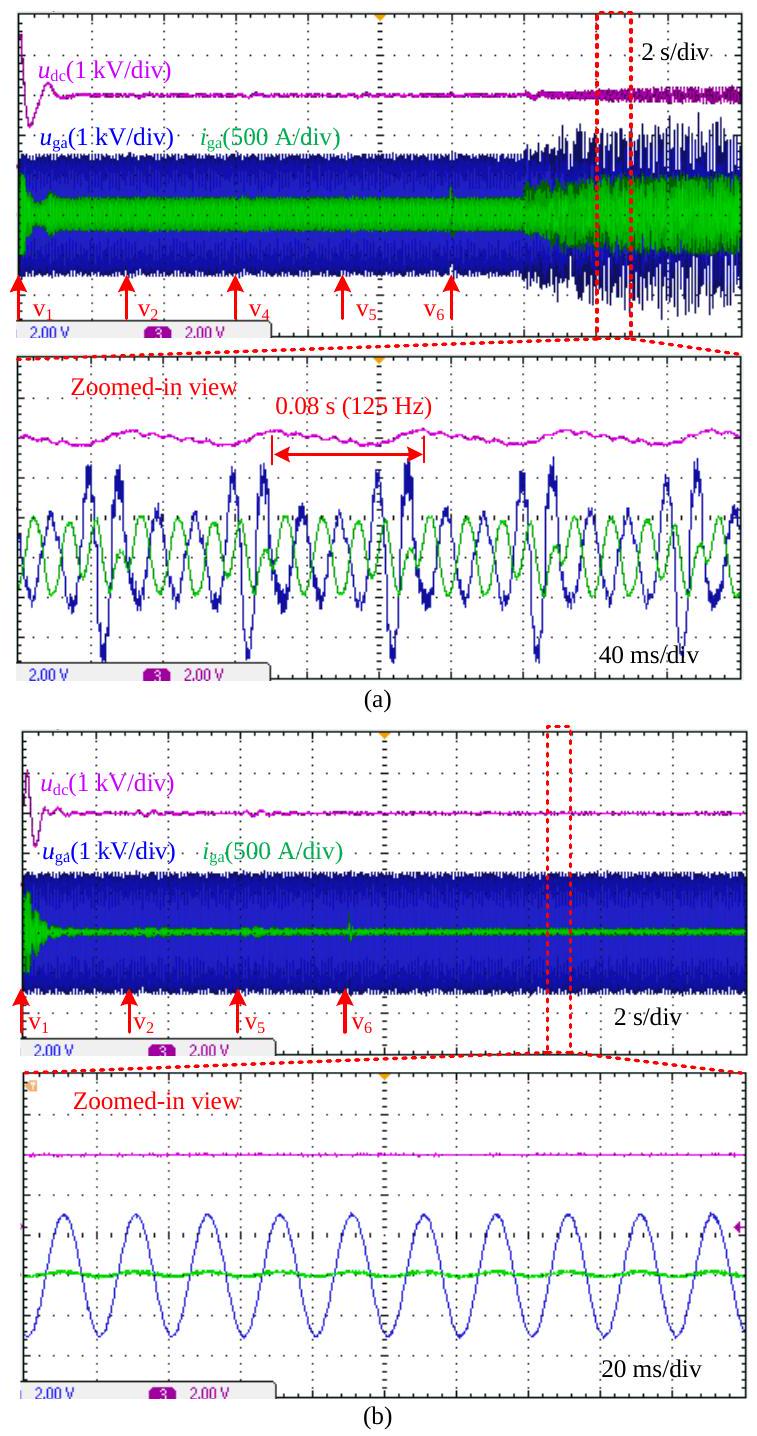}
\caption{Waveforms of line-side voltage and current, and dc-link voltage of vehicle. (a) Case 4. (b) Case 5.}
\end{center}
\end{figure}

\begin{enumerate}
  \setcounter{enumi}{1}
  \item With respect to railway system stability analysis, the CCM-based method incorporates the dynamic mobility of vehicles and their positional distribution, facilitating accurate assessments of both low-frequency and high-frequency instability issues within railway systems.
  \item By integrating the ResFNN network with the stability analysis method, the paper presents a framework for analyzing the stability of railway vehicle-grid systems, particularly under a double-sided feeding power supply mode.
\end{enumerate}

The effectiveness and advantages of the proposed methods are validated through experimental results derived from real operational data of actual railway lines.

\section*{Appendix}
\section*{A. Node admittance matrix of the vehicle-grid system}
According to the equivalent circuit of the vehicle network system shown in Fig. 10, based on the selected node voltages, the node admittance matrix of the vehicle network system is

\begin{table}[H]
\begin{center}
\caption{THE OPERATION POWER AND POSITION INFORMATION OF THE VEHICLE AT DIFFERENT TIMES}
\begin{tabular}{|l|l|l|l|l|l|l|}
\hline
\multirow{2}{*}{Time (h)} & \multicolumn{6}{|l|}{Operation power (kW), distance to traction substation A (kM)} \\
\hline
 & $\mathrm{v}_{1}$ & $\mathrm{v}_{2}$ & $\mathrm{v}_{3}$ & $\mathrm{v}_{4}$ & $\mathrm{v}_{5}$ & $\mathrm{v}_{6}$ \\
\hline
9:30 & 0, 0 & 0, 0 & 15,49 & 0, 50 & 0, 50 & 0, 50 \\
\hline
9:31 & 0, 0 & 0, 0 & 30, 46 & 0, 50 & 0, 50 & 0, 50 \\
\hline
9:32 & 0, 0 & 0, 0 & 150, 43 & 0, 50 & 0, 50 & 0, 50 \\
\hline
9:33 & 10, 1.6 & 0, 0 & 390, 40 & 0, 50 & 0, 50 & 0, 50 \\
\hline
9:34 & 120, 3.3 & 0, 0 & 550, 37 & 0, 50 & 0, 50 & 0, 50 \\
\hline
9:35 & 380, 5 & 10, 2.25 & 700, 34 & 10, 48.3 & 0, 50 & 0, 50 \\
\hline
9:36 & 560, 6.6 & 20, 4.5 & 720, 31 & 10, 46.6 & 0, 50 & 0, 50 \\
\hline
9:37 & 56, 8.3 & 60, 6.7 & 720, 28 & 120, 45 & 0, 50 & 0, 50 \\
\hline
9:38 & 560, 11.6 & 200, 9 & 720, 25 & 240, 43.3 & 0, 50 & 0, 50 \\
\hline
9:39 & 560, 15 & 340, 11.2 & 720, 22 & 360, 40 & 0, 50 & 0, 50 \\
\hline
9:40 & 560, 18.3 & 460, 13.5 & 700, 19 & 480, 36.6 & 0, 50 & 0, 50 \\
\hline
9:41 & 560, 21.6 & 580, 14 & -640, 16 & 600, 33.3 & 10, 48.2 & 0, 50 \\
\hline
9:42 & 560, 25 & 700,17 & -520, 13 & 680, 30 & 10, 46.3 & 0, 50 \\
\hline
9:43 & 560, 28.3 & 720, 20 & -220, 10 & 720, 26.6 & 100, 44.5 & 0, 50 \\
\hline
9:44 & 560, 31.6 & 720, 23 & -100, 7 & 720, 23.3 & 250, 40.8 & 0, 50 \\
\hline
9:45 & 560, 35 & 720, 26 & -10, 4 & 720, 20 & 440, 37.2 & 0, 50 \\
\hline
9:46 & 560, 38.3 & 720, 29 & 10, 0 & 720, 16.7 & 560, 33.5 & 0, 50 \\
\hline
9:47 & 560, 41.6 & 720, 32 & 0, 0 & 550, 13.3 & 700, 29.8 & 10, 49.9 \\
\hline
9:48 & 560, 43.3 & 720, 35 & 0, 0 & -550, 10 & 720, 26.2 & 100, 46.2 \\
\hline
9:49 & 460, 45 & 280, 38 & 0, 0 & -430, 8.3 & 720, 22.5 & 250, 42.5 \\
\hline
9:50 & -700, 46.6 & -720, 41 & 0, 0 & -270, 6.7 & 720, 18.8 & 440, 38.8 \\
\hline
9:51 & -340, 48.3 & -520, 43 & 0, 0 & -160, 5 & 550, 15.1 & 560, 35.2 \\
\hline
9:52 & -100, 49.1 & -340, 45.2 & 0, 0 & -60, 3.3 & -720, 11.4 & 700, 31.5 \\
\hline
9:53 & 20,50 & -180, 47.5 & 0, 0 & 0, 0 & -500, 9.6 & 720, 27.8 \\
\hline
9:54 & 0, 50 & -50, 49.7 & 0, 0 & 0, 0 & -300, 7.8 & 720, 24.2 \\
\hline
9:55 & 0, 50 & -10, 50 & 0, 0 & 0, 0 & -150, 5.9 & 720, 20.5 \\
\hline
9:56 & 0, 50 & 10, 50 & 0, 0 & 0, 0 & -10, 4.1 & 720, 16.8 \\
\hline
9:57 & 0, 50 & 10, 50 & 0, 0 & 0, 0 & 10, 0 & -300, 13.1 \\
\hline
9:58 & 0, 50 & 0, 50 & 0, 0 & 0, 0 & 0, 0 & -150, 9.5 \\
\hline
9:59 & 0, 50 & 0, 50 & 0, 0 & 0, 0 & 0, 0 & -10, 5.8 \\
\hline
10:00 & 0, 50 & 0, 50 & 0, 0 & 0, 0 & 0, 0 & 10, 2.2 \\
\hline
\end{tabular}
\end{center}
\end{table}

shown in (11). Due to space limitation, the superscript 'total' is omitted.

\section*{B. Vehicle operation information}
The vehicle operation power and location information from 9:30 to 10:00 are provided in Table IV.

\begin{equation*}
\resizebox{0.98\textwidth}{!}{$\displaystyle
\boldsymbol{Y}_{\text{sys }}=
{\left[\begin{array}{cccccccc}
Y_{n, 1}+Y_{1, m} & & & & -Y_{1, m} & -Y_{n, 2} & -Y_{2, m} & \\
& Y_{n, 2}+Y_{2, m} & & & -Y_{n, 3} & -Y_{3, m} & -Y_{n, 4} & -Y_{4, m} \\
& & Y_{n, 3}+Y_{3, m} & & Y_{n, 4}+Y_{4, m} & & & \\
-Y_{n, 1} & & -Y_{n, 3} & & Y_{n, 1}+Y_{n, 3}+2 Y_{n, m} & & & \\
-Y_{1, m} & & -Y_{3, m} & & & Y_{1, m}+Y_{3, m}+2 Y_{n, m} & -2 Y_{n, m} & \\
& -Y_{n, 2} & & -Y_{n, 4} & & -2 Y_{n, m} & Y_{n, 2}+Y_{n, 4}+2 Y_{n, m} & \\
& -Y_{2, m} & & -Y_{4, m} & -2 Y_{n, m} & & Y_{2, m}+Y_{4, m}+2 Y_{n, m} & \\
& & & & & & -2 Y_{n, m} & Y_{\mathrm{ga}}+2 Y_{n, m}
\end{array}\right.}
$}
\tag{11}
\end{equation*}

\section*{References}
[1] R. Hill, "Electric Railway Traction. Part 3. Traction Power Supplies," Power Eng. J., vol. 8, no. 6, pp. 275-286, Dec. 1994.\\[0pt]
[2] Y. Deng, Z. Liu, Y. Wang, and K. Huang, "Study on Application of All-Parallel DN Power Supply Mode on Montanic Electrified Railway," in Proc. 2015 Int. Conf. Electr. Inf. Technol. Rail Transp. Springer, Mar. 2016, pp. 153-162.\\[0pt]
[3] H. Wang, W. Mingli, and J. Sun, "Analysis of Low-Frequency Oscillation in Electric Railways Based on Small-Signal Modeling of VehicleGrid System in DQ Frame," IEEE Trans. Power Electron., vol. 30, no. 9, pp. 5318-5330, Sep. 2015.\\[0pt]
[4] K. Song, W. Mingli, S. Yang, Q. Liu, V. G. Agelidis, and G. Konstantinou, "High-order Harmonic Resonances in Traction Power Supplies: A Review Based on Railway Operational Data, Measurements, and Experience," IEEE Trans. Power Electron., vol. 35, no. 3, pp. 25012518, Mar. 2019.

[5] Railway Applications-Fixed Installations and Rolling Stock-Technical Criteria for the Coordination Between Electric Traction Power Supply Systems and Rolling Stock to Achieve Interoperability, CENELEC Standard EN 50388-1, 2022.\\[0pt]
[6] H. Zhang, Z. Liu, S. Wu, and Z. Li, "Input Impedance Modeling and Verification of Single-Phase Voltage Source Converters Based on Harmonic Linearization," IEEE Trans. Power Electron., vol. 34, no. 9, pp. 8544-8554, Sep. 2018.\\[0pt]
[7] D. Xie, C. Lin, Q. Deng, H. Lin, C. Cai, T. Basler, and X. Ge, "Simple Vector Calculation and Constraint-Based Fault-Tolerant Control for a Single-Phase CHBMC," IEEE Trans. Power Electron., 2024.\\[0pt]
[8] R. Pintelon and J. Schoukens, System Identification: A Frequency Domain Approach. 2nd ed. Hoboken, NJ, USA: Wiley, 2012.\\[0pt]
[9] G. Rojas-Dueñas, J.-R. Riba, and M. Moreno-Eguilaz, "Nonlinear Least Squares Optimization for Parametric Identification of DC-DC Converters," IEEE Trans. Power Electron., vol. 36, no. 1, pp. 654-661, Jan. 2020.\\[0pt]
[10] F. Hafiz, A. Swain, E. M. Mendes, and L. A. Aguirre, "Multiobjective Evolutionary Approach to Grey-Box Identification of Buck Converter," IEEE Trans. Circuits Syst. I, vol. 67, no. 6, pp. 2016-2028, Jun. 2020.\\[0pt]
[11] V. Valdivia, A. Barrado, A. LÁzaro, P. Zumel, C. Raga, and C. FernÁndez, "Simple Modeling and Identification Procedures for "Black-Box" Behavioral Modeling of Power Converters Based on Transient Response Analysis," IEEE Trans. Power Electron., vol. 24, no. 12, pp. 2776-2790, Dec. 2009.\\[0pt]
[12] V. Valdivia, A. Barrado, A. Lazaro, M. Sanz, D. L. del Moral, and C. Raga, "Black-Box Behavioral Modeling and Identification of DC-DC Converters with Input Current Control for Fuel Cell Power Conditioning," IEEE Trans. Ind. Electron., vol. 61, no. 4, pp. 1891-1903, Apr. 2013.\\[0pt]
[13] A. Frances-Roger, A. Anvari-Moghaddam, E. Rodriguez-Diaz, J. C. Vasquez, J. M. Guerrero, and J. Uceda, "Dynamic Assessment of COTS Converters-Based DC Integrated Power Systems in Electric Ships," IEEE Trans. Ind. Informat., vol. 14, no. 12, pp. 5518-5529, Dec. 2018.\\[0pt]
[14] P. Xiao, G. K. Venayagamoorthy, K. A. Corzine, and J. Huang, "Recurrent Neural Networks Based Impedance Measurement Technique for Power Electronic Systems," IEEE Trans. Power Electron., vol. 25, no. 2, pp. 382-390, Feb. 2009.\\[0pt]
[15] M. Zhang, X. Wang, D. Yang, and M. G. Christensen, "Artificial Neural Network Based Identification of Multi-Operating-Point Impedance Model," IEEE Trans. Power Electron., vol. 36, no. 2, pp. 1231-1235, Feb. 2020.\\[0pt]
[16] Y. Liao, Y. Li, M. Chen, L. Nordström, X. Wang, P. Mittal, and H. V. Poor, "Neural Network Design for Impedance Modeling of Power Electronic Systems Based on Latent Features," IEEE Trans. Neural Netw. Learn. Syst., vol. 35, no. 5, pp. 5968-5980, May 2024.\\[0pt]
[17] S. Zhao, Y. Peng, Y. Zhang, and H. Wang, "Parameter Estimation of Power Electronic Converters with Physics-Informed Machine Learning," IEEE Trans. Power Electron., vol. 37, no. 10, pp. 11567-11578, Oct. 2022.\\[0pt]
[18] M. Zhang, Q. Xu, and X. Wang, "Physics-Informed Neural Network Based Online Impedance Identification of Voltage Source Converters," IEEE Trans. Ind. Electron., vol. 70, no. 4, pp. 3717-3728, Apr. 2022.\\[0pt]
[19] Y. Liao, Z. Liu, G. Zhang, and C. Xiang, "Vehicle-Grid System Modeling and Stability Analysis with Forbidden Region-Based Criterion," IEEE Trans. Power Electron., vol. 32, no. 5, pp. 3499-3512, May 2016.\\[0pt]
[20] W. Cao, Y. Ma, F. Wang, L. M. Tolbert, and Y. Xue, "Low-Frequency Stability Analysis of Inverter-Based Islanded Multiple-Bus AC Microgrids Based on Terminal Characteristics," IEEE Trans. Smart Grid, vol. 11, no. 5, pp. 3662-3676, Sep. 2020.\\[0pt]
[21] H. Tao, H. Hu, X. Wang, F. Blaabjerg, and Z. He, "Impedance-Based Harmonic Instability Assessment in a Multiple Electric Trains and Traction Network Interaction System," IEEE Trans. Ind. Appl., vol. 54, no. 5, pp. 5083-5096, Sep. 2018.\\[0pt]
[22] X. Meng, Q. Zhang, Z. Liu, G. Hu, F. Liu, and G. Zhang, "Multiple Vehicles and Traction Network Interaction System Stability Analysis and Oscillation Responsibility Identification," IEEE Trans. Power Electron., vol. 39, no. 5, pp. 6148-6162, May 2024.\\[0pt]
[23] S. M. Lundberg and S.-I. Lee, "A Unified Approach to Interpreting Model Predictions," Proc. NeurIPS, vol. 30, Dec. 2017.\\[0pt]
[24] H. Wang, Z. Han, Z. Liu, and Y. Wu, "Deep Reinforcement Learning Based Active Pantograph Control Strategy in High-Speed Railway," IEEE Trans. Veh. Technol., vol. 72, no. 1, pp. 227-238, Sep. 2022.\\[0pt]
[25] H. Wang, Z. Liu, G. Hu, X. Wang, and Z. Han, "Offline MetaReinforcement Learning for Active Pantograph Control in High-Speed Railways," IEEE Trans. Ind. Informat., pp. 1-11, May 2024.\\[0pt]
[26] U. Faigle and W. Kern, "The Shapley Value for Cooperative Games Under Precedence Constraints," Int. J. Game Theory, vol. 21, pp. 249266, Sep. 1992.\\[0pt]
[27] Y. Liao, Z. Liu, H. Zhang, and B. Wen, "Low-Frequency Stability Analysis of Single-Phase System With DQ-Frame Impedance Approach-Part I: Impedance Modeling and Verification," IEEE Trans. Ind. Appl., vol. 54, no. 5, pp. 4999-5011, Sep. 2018.\\[0pt]
[28] F. Alonge, F. D'Ippolito, F. M. Raimondi, and S. Tumminaro, "Nonlinear Modeling of DC/DC Converters Using the Hammerstein's Approach," IEEE Trans. Power Electron., vol. 22, no. 4, pp. 1210-1221, Jul. 2007.\\[0pt]
[29] W. Zhou, R. E. Torres-Olguin, Y. Wang, and Z. Chen, "A GrayBox Hierarchical Oscillatory Instability Source Identification Method of Multiple-Inverter-Fed Power Systems," IEEE Trans. Emerg. Sel. Topics Power Electron., vol. 9, no. 3, pp. 3095-3113, Jun. 2020.\\[0pt]
[30] L. Fan, Z. Miao, P. Koralewicz, S. Shah, and V. Gevorgian, "Identifying DQ-domain Admittance Models of a 2.3 -MVA Commercial GridFollowing Inverter via Frequency-Domain and Time-Domain Data," IEEE Trans. Energy Convers., vol. 36, no. 3, pp. 2463-2472, Sep. 2020.\\[0pt]
[31] L. Luo and S. V. Dhople, "Spatiotemporal Model Reduction of InverterBased Islanded Microgrids," IEEE Trans. Energy Convers., vol. 29, no. 4, pp. 823-832, Dec. 2014.\\[0pt]
[32] T. Ho, Y. Chi, J. Wang, K. Leung, L. Siu, and C. Tse, "Probabilistic Load Flow in AC Electrified Railways," IEE Proc. Elect. Power Appl., vol. 152, no. 4, pp. 1003-1013, Jul. 2005.\\[0pt]
[33] B. Wen, D. Boroyevich, R. Burgos, P. Mattavelli, and Z. Shen, "Analysis of DQ Small-Signal Impedance of Grid-Tied Inverters," IEEE Trans. Power Electron., vol. 31, no. 1, pp. 675-687, Jan. 2015.\\[0pt]
[34] C. Desoer and Y.-T. Wang, "On the Generalized Nyquist Stability Criterion," IEEE Trans. Autom. Control, vol. 25, no. 2, pp. 187-196, Apr. 1980.\\[0pt]
[35] H. Lin, H. S.-H. Chung, R. Shen, and Y. Xiang, "Enhancing Stability of DC Cascaded Systems with CPLs Using MPC Combined with NI and Accounting for Parameter Uncertainties," IEEE Trans. Power Electron., 2024.\\[0pt]
[36] B. Lu, Y. Song, Z. Liu, G. Tao, X. Wang, Q. Zhang, and Z. Li, "Evolution Analysis of Wheel Polygon Wear Considering the Effect of Interharmonics in Electrical Traction Drive System," Mech. Mach. Theory, vol. 191, p. 105470, Jan. 2024.

\end{document}